\documentclass[conference]{IEEEtran}
\IEEEoverridecommandlockouts

\usepackage{cite}
\usepackage{amsmath,amssymb,amsfonts}
\usepackage{graphicx}
\usepackage{textcomp}
\usepackage{xcolor}
\usepackage{url}
\usepackage{booktabs}
\usepackage{subfig}
\usepackage{comment}
\usepackage{multirow}

\def\BibTeX{{\rm B\kern-.05em{\sc i\kern-.025em b}\kern-.08em
    T\kern-.1667em\lower.7ex\hbox{E}\kern-.125emX}}

\begin{document}

\title{(POSTER) From Sensors to Insight: Rapid, Edge-to-Core Application Development for Sensor-Driven Applications}

\author{\IEEEauthorblockN{
Komal Thareja\IEEEauthorrefmark{1},
Anirban Mandal\IEEEauthorrefmark{1},
Ewa Deelman\IEEEauthorrefmark{3}
}

\IEEEauthorblockA{\IEEEauthorrefmark{1}Renaissance Computing Institute, University of North Carolina Chapel Hill, NC, USA}

\IEEEauthorblockA{\IEEEauthorrefmark{3}Information Sciences Institute, University of Southern California, Marina del Rey, CA, USA}
}

\maketitle

\begin{abstract}
Scientists increasingly rely on sensor-based data; however transforming raw streams into insights across the edge-to-cloud continuum remains difficult due to the breadth of expertise required to coordinate the necessary data and computation flow. This paper introduces a pattern-based, AI-assisted methodology for rapid development of sensor-driven applications. Using Pegasus workflows executing on the FABRIC testbed, we demonstrate a 5-step development loop that shifts workflow construction and deployment from code-first to intent-first design. Starting from an existing Orcasound hydrophone workflow as a reusable template, we generate and refine workflows for air quality, earthquake, and soil moisture monitoring applications. We further show how these workflows extend to edge resources---including BlueField-3 DPUs and Raspberry Pis---through configuration and placement rather than workflow redesign. Our evaluation, from the perspective of a novice Pegasus user, shows that AI-assisted pattern reuse compresses multi-stage workflow development to 1--1.5 days per workflow while preserving the rigor and portability of workflow-based execution.
\end{abstract}


\section{Introduction and Motivation}

Sensor-driven applications are a cornerstone of environmental monitoring, smart infrastructure, and cyber-physical systems. These applications require the execution of end-to-end processing pipelines that transform raw sensor data into actionable insight across the edge-to-cloud continuum. However, moving from a scientific question (e.g., ``How does local air quality impact respiratory health trends?'') to a production workflow requires deep infrastructure expertise---what we term the \textbf{``blank page bottleneck.''} Scientists often spend more time on execution mechanics (workflow formats, container orchestration, infrastructure provisioning) than exploring and analyzing their data.

Workflow management systems like Pegasus~\cite{deelman2015pegasus} and programmable infrastructures like FABRIC~\cite{baldin2019fabric} offer powerful abstractions for managing distributed applications, including emerging hardware accelerators such as Data Processing Units (DPUs)~\cite{bluefield2022}. However, effectively leveraging these tools requires understanding configuration models, deployment options, and infrastructure interfaces. Meanwhile, AI code assistants~\cite{chen2021evaluating} show promise for boilerplate code generation but produce fragile, standalone scripts when unconstrained, limiting portability, provenance tracking, and fault tolerance.

We address this gap through \textbf{pattern-based engineering with AI assistance}: users provide an AI assistant (e.g., Claude~\cite{anthropic_claude}) with an existing Pegasus workflow pattern and describe a new use case in natural language. The AI adapts the pattern---retaining dependency logic while swapping domain-specific components---producing a working first draft. Grounding AI in validated patterns prevents hallucinations and ensures that the generated code follows established best practices.

The paper makes the following {\bf contributions}: (1)~a pattern-based methodology and 5-step development loop for {\it intent-first} workflow construction; (2)~four case studies (air quality, earthquake, soil moisture, Orcasound) deployed across edge-to-core continuum including Edge DPUs (e.g. BlueField-3); (3)~a user-centric evaluation demonstrating that AI-assisted pattern reuse compresses development from weeks to days for novice users; (4)~practical insights on why Pegasus workflows are synergistic with LLM-assisted generation due to their explicit, readable, composable, and execution-agnostic nature.

\section{Methodology}

\begin{figure}[t]
  \centering
  \includegraphics[width=0.95\linewidth]{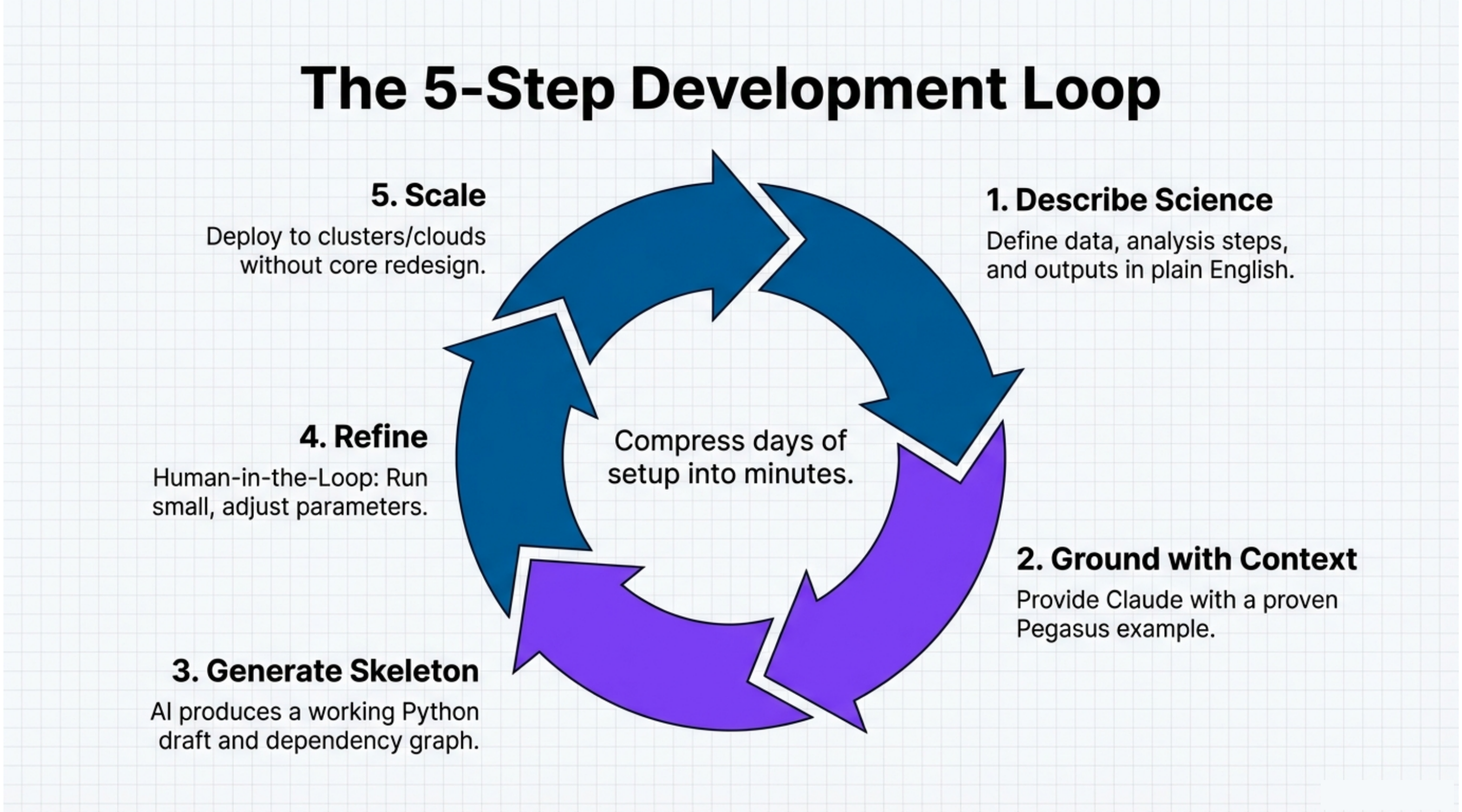}
  \caption{Pattern-based, AI-assisted 5-step development loop.}
  \vspace{-15pt}
  \label{fig:five_step_loop}
\end{figure}

Our methodology combines \textit{pattern-based workflow engineering} with \textit{AI-assisted development}, guided by four design principles: \textbf{(1)~Intent-first development}---scientists express data sources, analysis steps, and expected outputs in natural language rather than low-level execution mechanics; \textbf{(2)~Pattern reuse over greenfield design}---new applications are derived by adapting existing, validated workflow patterns; \textbf{(3)~Separation of concerns}---workflow logic, execution placement, and infrastructure provisioning are explicitly separated; \textbf{(4)~Human-in-the-loop AI assistance}---AI tools assist with workflow generation and debugging but do not replace workflow engines or schedulers. Figure~\ref{fig:five_step_loop} summarizes the approach.

Our methodology is structured around a {\bf five-step development loop}:
\textbf{(1)~Describe Science}---define data sources, analysis steps, and outputs in plain English;
\textbf{(2)~Ground with Context}---provide the AI with a proven Pegasus workflow as a structural template;
\textbf{(3)~Generate Skeleton}---AI produces executable Python code with task definitions, I/O declarations, and dependency graphs;
\textbf{(4)~Refine}---human-guided iteration to adjust resource requirements, containerization, and parallelization;
\textbf{(5)~Scale}---deploy to edge, clusters, or clouds without workflow redesign using Pegasus catalogs and HTCondor~\cite{thain2005distributed} placement.

All applications are represented as Pegasus Directed Acyclic Graphs (DAGs) with explicit data dependencies. This representation is particularly compatible with AI-assisted development because it is \textit{explicit}, \textit{readable}, \textit{composable}, and \textit{execution-agnostic}. It is also and open source project with documentation available online for over 20 years, making commercial LLMs work out of the box. For edge-to-cloud workflows, HTCondor ClassAds steer tasks to appropriate nodes (e.g., DPU-enabled edge workers vs.\ CPU-only cloud workers). Infrastructure is provisioned programmatically using FABlib on the FABRIC testbed, with typical deployments including a submit node hosting Pegasus/HTCondor, edge workers near data sources, and cloud workers for compute-intensive stages.

\section{Case Studies}

We demonstrate progressive workflow reuse through a chaining pattern: Orcasound $\rightarrow$ air quality $\rightarrow$ earthquake and soil moisture. Each workflow inherits validated structure from its predecessor, compounding productivity gains across successive applications.

\subsection{Orcasound: Baseline Pattern}

The Orcasound hydrophone workflow~\cite{orcasound2020} served as the foundational pattern, implementing a canonical sensor-driven structure: \textit{ingest} $\rightarrow$ \textit{preprocess} $\rightarrow$ \textit{transform} $\rightarrow$ \textit{infer} $\rightarrow$ \textit{aggregate}. Concretely, it ingests audio segments, converts them into analysis-ready formats, performs ML inference to detect Orca vocalizations, and aggregates predictions across sensors and time windows. Its explicit dependency graph, containerized transformations, and separation of workflow logic from execution placement make it ideal for cross-domain adaptation.

\subsection{Air Quality: Incremental Evolution}

\begin{figure}[t]
  \centering
  \includegraphics[width=0.95\linewidth]{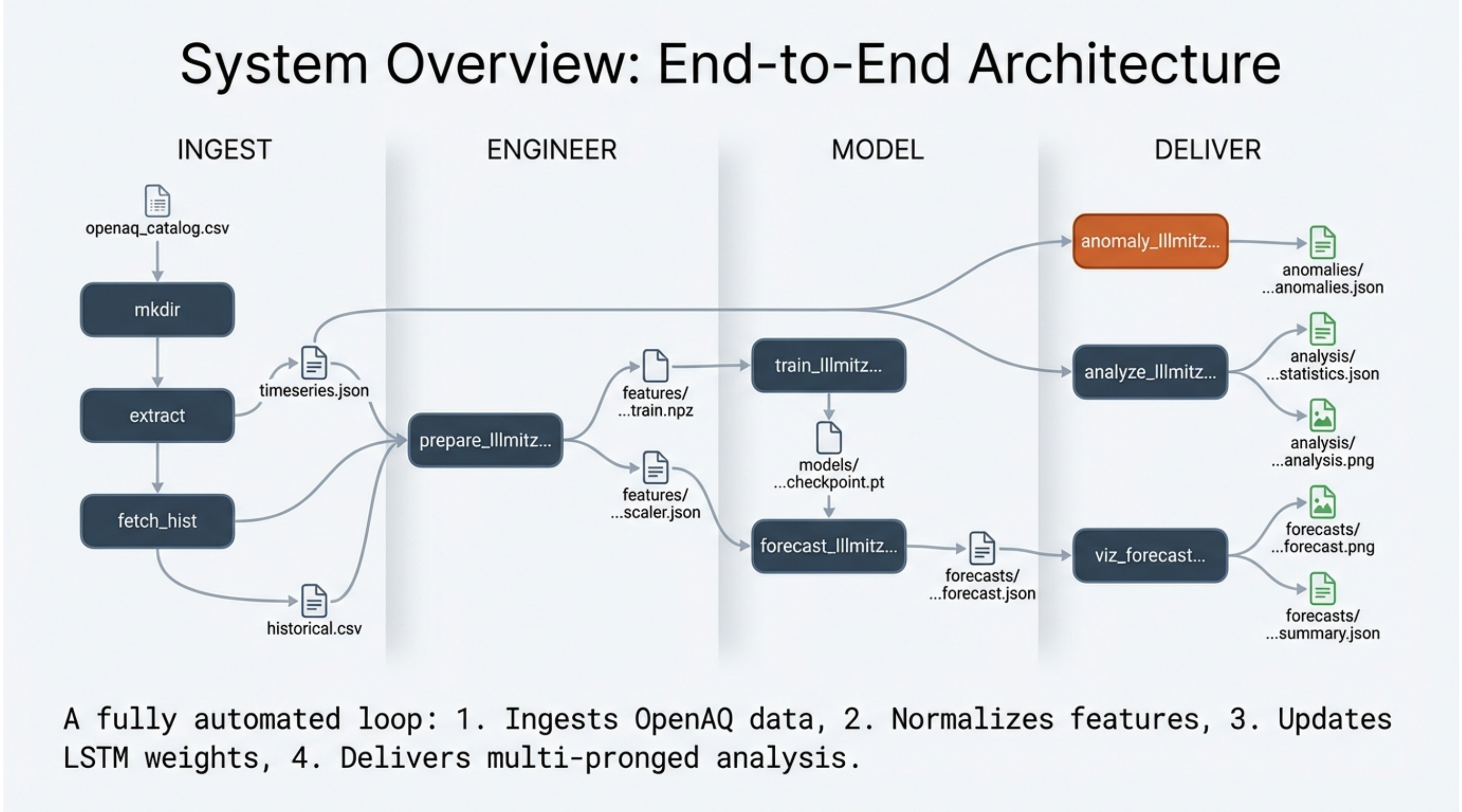}
  \caption{Air quality workflow DAG.}
  \vspace{-15pt}
  \label{fig:aq-dag}
\end{figure}

\begin{figure}[t]
  \centering
  \includegraphics[width=0.95\linewidth]{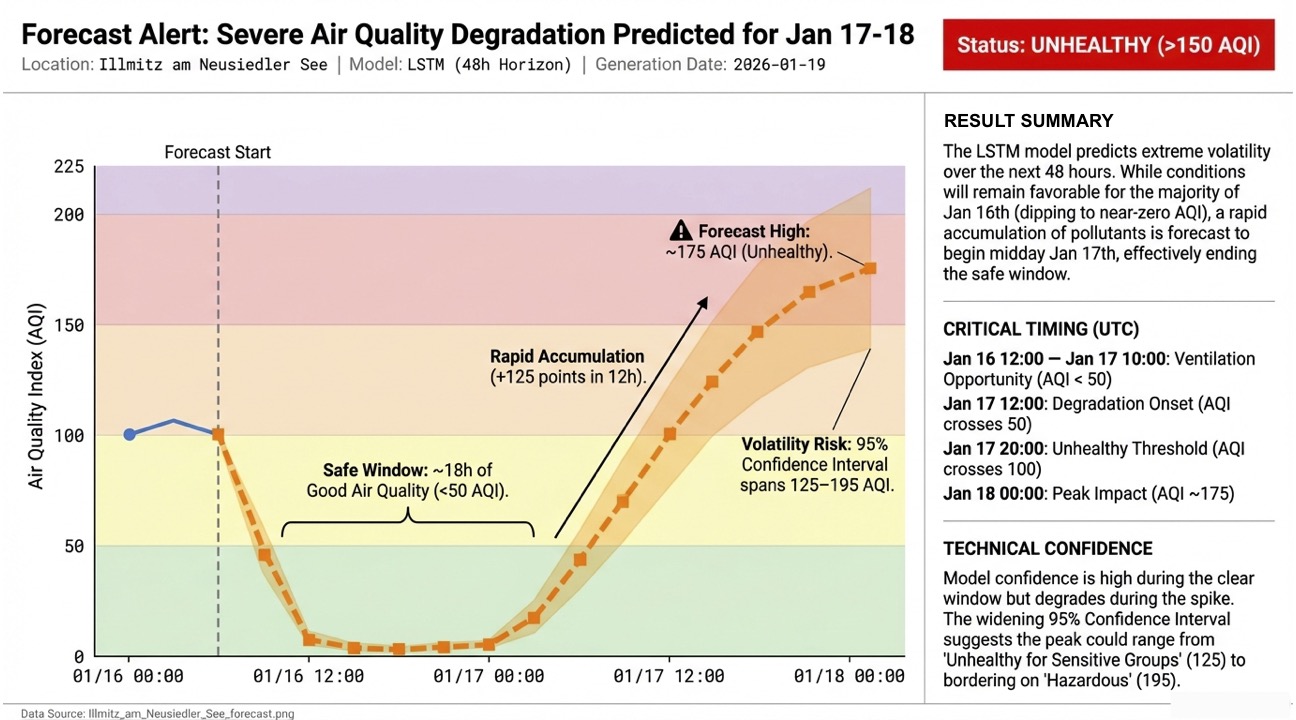}
  \caption{Representative AQI forecast and anomaly detection.}
  \vspace{-15pt}
  \label{fig:aq-result}
\end{figure}

Starting from Orcasound, the air quality workflow~\cite{airquality-workflow} was built incrementally: (i)~a baseline ingest--analyze--detect pipeline using OpenAQ~\cite{openaq} data (using 3--4 AI prompts), (ii)~extension to LSTM-based ML forecasting (3--4 additional prompts), and (iii)~integration of SAGE~\cite{sage2021} sensor network data (2--3 prompts). Each extension preserved upstream logic. Early executions exposed realistic runtime errors---e.g., missing API credentials inside execution containers causing HTCondor jobs to enter \texttt{Held} state---which were diagnosed through AI-assisted log analysis across Pegasus, HTCondor, and container layers. Figure~\ref{fig:aq-dag} shows the resulting workflow DAG and Figure~\ref{fig:aq-result} shows a representative AQI forecast output.

\subsection{Earthquake and Soil Moisture}

Both workflows reused the validated air quality pipeline as their immediate template. The earthquake workflow~\cite{earthquake-workflow} replaced ingestion with USGS~\cite{usgs} seismic feeds and introduced conditional branching to trigger aftershock analysis, pattern detection, and hazard assessment for high-magnitude events. The soil moisture workflow~\cite{soilmoisture-workflow} adapted ingestion for in-situ soil sensors and weather data, followed by feature joining and regression-based prediction. In both cases, the underlying execution model and deployment configuration remained largely unchanged.

\subsection{Edge Integration: DPUs and Raspberry Pi}

After validating cloud-based executions, we extended workflows to incorporate edge resources \textit{without modifying the abstract workflow}. For the soil moisture workflow, I/O-intensive preprocessing stages were offloaded to BlueField-3 DPUs via Pegasus site mappings, enabling experimentation with alternative placements along the edge-to-cloud continuum. A Raspberry Pi was also configured as an edge participant for data ingestion and lightweight preprocessing, with downstream analytics executing on cloud resources. Both cases demonstrate that heterogeneous edge devices can become first-class participants in end-to-end Pegasus workflows using standard scheduling and execution mechanisms---no bespoke edge orchestration layers are required.

\section{Evaluation}

Our evaluation targets \textit{user productivity} from the perspective of a novice Pegasus user whose prior experience was limited to a single tutorial~\cite{ACCESS_Pegasus}. We used Claude~\cite{anthropic_claude} (Opus 4.5) as the primary AI assistant, supplemented by ChatGPT~\cite{openai_codex} and Gemini~\cite{google_gemini} during debugging sessions. All workflows were deployed on the FABRIC testbed using FABlib for provisioning and FABRIC notebooks~\cite{fabric_pegasus_artifact} as contextual grounding.

\noindent\textbf{Key results.}
Table~\ref{tab:effort} summarizes the development effort.

\begin{table}[t]
\centering
\caption{Development effort per workflow stage.}
\vspace{-5pt}
\label{tab:effort}
\footnotesize
\begin{tabular}{lccc}
\toprule
\textbf{Workflow / Stage} & \textbf{Prompts} & \textbf{Exec. Attempts} & \textbf{Dev. Time} \\
\midrule
AQ: Baseline & 3--4 & 7 & 0.5--1 day \\
AQ: ML extension & 3--4 & 2--3 & $\sim$0.5 day \\
AQ: SAGE data & 2--3 & 2--3 & $\sim$0.25 day \\
\midrule
Earthquake & similar & fewer & $\sim$1 day \\
Soil moisture & similar & fewer & $\sim$1 day \\
\midrule
\multicolumn{3}{l}{\textbf{Total (all workflows)}} & \textbf{4--5 days} \\
\bottomrule
\end{tabular}
\vspace{-10pt}
\end{table}

\noindent\textbf{Cross-Workflow Observations.}
Across all workflows, we observed consistent trends:
\begin{itemize}
    \item \textbf{Rapid time-to-first-workflow}: Runnable workflows were generated within 3--4 prompts when grounded in an existing pattern.
    \item \textbf{Decreasing iteration cost}: Later workflows required fewer execution attempts due to inherited configurations and prior debugging experience.
    \item \textbf{Debugging dominates effort}: Most iteration effort was spent resolving execution-time configuration issues (credential propagation, data staging, container runtime) rather than correcting workflow logic.
    \item \textbf{Compounding productivity gains}: Progressive reuse (Orcasound $\rightarrow$ AQ $\rightarrow$ earthquake/soil moisture) reduced development effort for each successive workflow.
\end{itemize}
The total development effort across all workflows was approximately 4--5 days (24--40 person-hours), or 8--12 hours per workflow---a substantial reduction for a novice user who would otherwise need weeks to learn workflow abstractions, configure execution environments, and diagnose distributed runtime failures.

\section{Conclusions and Future Work}

We presented an experience-driven methodology combining pattern-based workflow engineering with AI-assisted development for sensor-driven, edge-to-core applications. Using Pegasus on the FABRIC testbed, we demonstrated how validated workflow patterns can be incrementally adapted across domains---air quality analysis, earthquake processing, soil moisture prediction, and hydrophone-based audio analytics---and execution environments, including BlueField-3 DPUs and Raspberry Pis, without redesigning the abstract workflow.

Our evaluation highlights that AI assistance is most effective when grounded in proven workflow patterns and applied to iterative refinement and debugging. Progressive reuse compounds productivity gains: each successive workflow benefits from inherited structure, configuration, and debugging experience. This work has limitations, including an evaluation focus on development effort rather than application performance and reliance on a single user's experience.

Future directions include extending the methodology to team-based development effort evaluation and longer-lived workflows, incorporating execution feedback for automated workflow adaptation, comparative performance evaluation against manually constructed workflows by experts, integrating AI agents that proactively suggest optimizations and placement strategies, and leveraging tools like Kiso~\cite{Kiso2026} for automated infrastructure provisioning across the edge-to-cloud continuum.

\section*{Acknowledgments}
This work is supported by the US National Science Foundation grants \#2403051 and \#2513101. We acknowledge SAGE (NSF award \#2436842 and FABRIC Testbed (NSF \#2330891).

\bibliographystyle{IEEEtran}
\bibliography{references}

\end{document}